\documentclass[final, runningheads]{llncs}

\usepackage{amsmath}
\usepackage{amssymb}
\usepackage{comment}
\usepackage{fancyhdr}
\usepackage[T1]{fontenc}
\usepackage{graphicx}
\usepackage{mathtools}
\usepackage{multirow}
\usepackage{stmaryrd}
\usepackage{xcolor}

\newcommand{\methodName}{T-SAIRUS}

\newcommand{\itadata}{\footnotesize \textsl{ITADATA2024: The 3$^{\text{rd}}$ Italian Conference on Big Data and Data Science}}

\fancypagestyle{empty}{\fancyhf{}\fancyhead[C]{\itadata}
  \fancyhead[R]{\includegraphics[width=.05\textwidth]{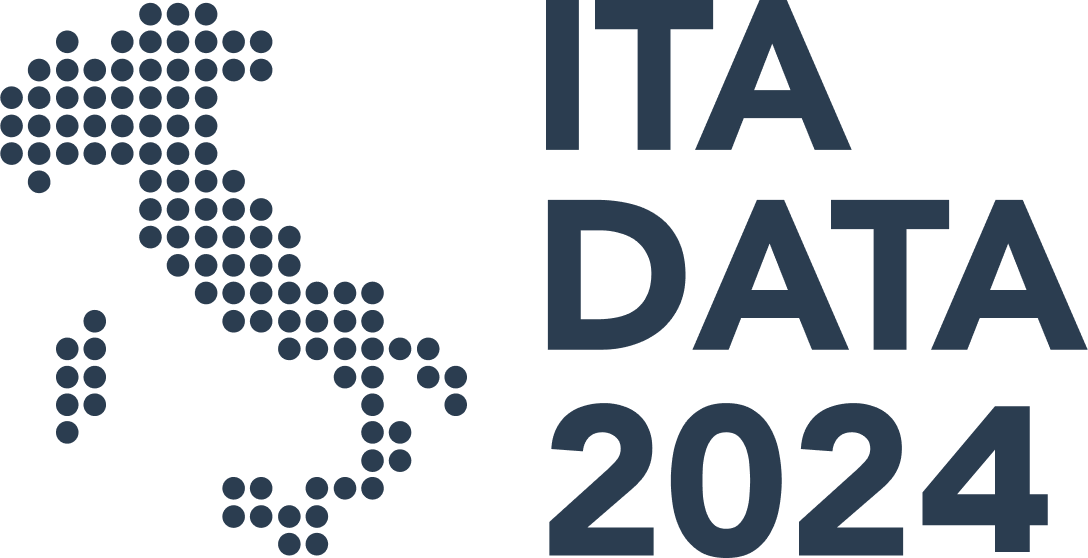}}}

\makeatletter
\begin{document}

\title{Multi-view Learning for the Identification of Risky Users in Dynamic Social Networks}

\author{Francesco Benedetti\inst{1,2,3}\orcidID{0009-0009-6610-9846} \and
Antonio Pellicani\inst{1,3}\orcidID{0000-0002-4193-3486} \and
Gianvito Pio\inst{1,3}\orcidID{0000-0003-2520-3616} \and
Michelangelo Ceci\inst{1,3,4}\orcidID{0000-0002-6690-7583}} 

\authorrunning{F. Benedetti et al.}

\institute{Dept. of Computer Science, University of Bari, Bari, Italy \and
Dept. of Computer Science, University of Pisa, Pisa, Italy \and Data Science Lab, National Interuniversity Consortium for Informatics, Rome, Italy \and Dept. of Knowledge Technologies, Jo\v{z}ef Stefan Institute, Ljubljana, Slovenia}

\maketitle             

\begin{abstract}
Technological progress in the last few decades has granted an increasing number of people access to social media platforms such as Facebook, X (formerly Twitter), and Instagram. Consequently, the potential risks associated with these services have also risen due to users exploiting these services for malicious purposes. The platforms have tools capable of detecting and blocking dangerous users, but they primarily focus on the content posted by users and usually overlook additional factors, such as the relationships among users. Another key aspect to consider is that users' beliefs and interests evolve over time. Therefore, a user who can be considered safe at one moment might later become malicious, and vice versa. This work describes a novel approach to node classification in temporal graphs, aimed at classifying users in social networks. The method was evaluated on a real-world scenario and was compared to a state-of-the-art system that treats the network as a static entity. Experiments showed that taking into account the temporal evolution of the network, in terms of node features and connections, is beneficial.

\end{abstract}

\section{Introduction}
\label{sec1}
Social media platforms play a crucial role in modern society, because they enable connections between users and their friends, relatives, or like-minded people, often overcoming physical distances.
Over the time, the number of users who steadily utilize these platforms has grown exponentially, making social media an essential communication channel. 
However, the power of these platforms can be exploited for malicious purposes, such as inciting hate speech against minority groups, propagating extremist ideologies, facilitating the growth of radical communities, and even recruiting vulnerable individuals into terrorist organizations.

Several reports show how radicals take advantage of social media \cite{bib1,thompson2011radicalization}, and the problem becomes more serious as the size of social media grows. 
Thus, detecting dangerous users on social media became essential not only for the interests of platforms, which aim to maintain a secure and trustworthy environment for their users, but also for safety reasons. However, this task is often impractical for humans, due to the enormous mass of data required to analyze.
To address this challenge, a possible solution relies on the use of techniques falling into Social Network Analysis (SNA), a discipline encompassing several fields ranging from sociology to mathematics, with the goal of gaining insights from social networks.
SNA techniques are used for a wide range of tasks, such as community detection \cite{com_detection,com_detection2}, spammer detection \cite{DBLP:conf/aaai/ZhuWZLLY12} and user classification, which is the task tackled in this work.
Practically, it involves learning a model able to assign a label to a user in the network, and in our specific case, the final goal is to determine whether the user is \textit{risky} or \textit{safe}.

It is preferable that the learnt model is aware of the main perspectives characterizing social media, specifically: \textit{i)} the semantics of the data generated by users, which indicates their beliefs and intentions, \textit{ii)} the topology of the social network, which allows understanding how users interact with each other, and \textit{iii)} the spatial distance among users, which allows considering clusters of spatially close users, who are subject to the same real-life events, share the same culture and are more likely to meet in person.

Even if some existing methods already consider these  perspectives \cite{corizzo2023huri,sairus}, they usually look at the social network as a static entity without taking into account its intrinsic high dynamicity. Indeed, user relationships, as well as the semantics of the posted contents and the spatial distance among users, are in continuous evolution. Ignoring the dynamic nature of the network causes the model to miss users who were previously safe but suddenly began to radicalize.
In this work, we aim to address this issue by proposing \methodName{}, a temporally-aware social network analysis tool that considers different temporal snapshots of the network and analyzes them separately. Inspired by \cite{sairus}, each temporal snapshot is analyzed by means of a stacked generalization of different models \cite{stacked}, each responsible for analyzing and integrating the information conveyed by the diverse perspectives of the social network, in a multi-view learning setting. Then, the predictions for all the snapshots are combined to obtain the final user label. In this way, we can consider the evolution of the user behavior.

Furthermore, existing systems that solve the task of risky user identification typically work in a semi-supervised transductive setting. In this setting, the model is trained on a specific network and can only make predictions for (unlabeled) users already present in such a network at training time. Consequently, when a new user is added to the network, these systems require a full retraining to recognize and classify the new user. While \methodName{} also operates in a transductive setting, it offers a key advantage: it requires a less complex training process. Indeed, when a new user is added, \methodName{} only needs  to be trained on the last temporal snapshot, which includes the newly added user, allowing it to recycle models previously learned for earlier snapshots.

The remainder of the paper is organized as follows: Section \ref{sec3} describes the details of the proposed strategy; Section \ref{sec4} presents the experimental setup and discusses the results of our experimental evaluation. Finally, Section \ref{sec:conclusions} draws some conclusions and outlines possible future work.

\section{The method \methodName{}}
\label{sec3}
Before introducing \methodName{}, we provide a formal definition of a social network $\mathcal{G}$ as a four-tuple $\langle N, C, E_C, E_T \rangle$, where:
\begin{itemize}
    \item $N = N_L \cup N_U$ is the set of nodes, each of which represents a user. Specifically, we indicate with $N_L$ the labeled users (\textit{safe} or \textit{risky}), and with $N_U$ the unlabeled users, with $N_L \cap N_U = \emptyset$.
    \item $C$ are the posts. Each post consists of text, and is associated with a timestamp and a geographical location.
    \item $E_C \subseteq N \times C$ is a set of links between users and textual content. Each link depicts the action of writing and posting a textual content.
    \item $E_T \subseteq N \times N$ defines the topology of the social network, where each connection represents a specific social relationship, e.g. \textit{like} or  \textit{follow}.

\end{itemize}

\noindent In order to properly take into account the temporal aspect, we apply the snapshot-based strategy presented in \cite{temporal_overview}. Specifically, we use the timestamps associated with the posts to split the network into a series of $T$ consecutive, partially overlapping, snapshots, $\langle \mathcal{G}^1, \mathcal{G}^2, ..., \mathcal{G}^T \rangle$.

We will refer to the $i$-th snapshot as $\mathcal{G}^{(i)} = \langle N^{(i)}, C^{(i)}, {E_C}^{(i)}, {E_T}^{(i)} \rangle$.
We use the first $T-1$ snapshots for training the model to be used for predictions of nodes in the last snapshot (and the next ones). 
Since our approach works in a \textit{within-network} setting \cite{desrosiers2009within}, it requires the unlabeled users in the last snapshot to appear in at least one of the previous snapshots.

\begin{figure}[tb]
    \centering    \includegraphics[width=0.8\textwidth]{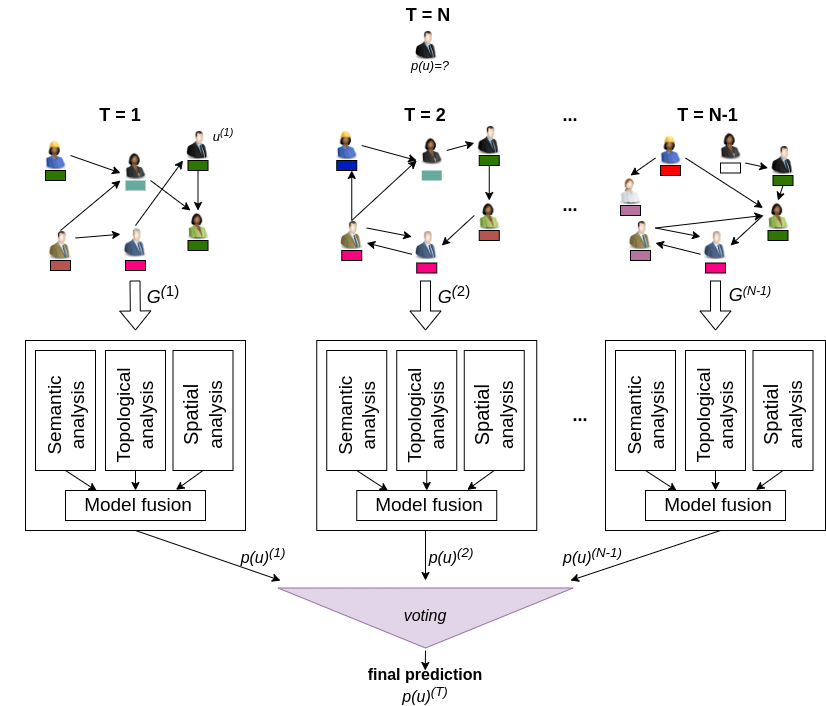}
    \caption{T-SAIRUS framework. The figure emphasizes how the network relationships change over time, while the coloured bars below each user represent the score associated with them, which can also evolve.}
    \label{fig:diagram}
\end{figure}

In Figure \ref{fig:diagram}, the structure of the proposed framework is highlighted.
For each temporal snapshot $\mathcal{G}^1, \mathcal{G}^2, ..., \mathcal{G}^T$, the framework follows the stacked generalization approach and performs four key steps: \textit{i)} semantic analysis of the posted content, \textit{ii)} topological analysis of the relationship network, \textit{iii)} spatial analysis of the users' geographic proximity, and \textit{iv)} model fusion. Finally, the predictions obtained from each temporal snapshot are combined through a voting mechanism. In the following subsections, we will briefly explain each of these components.

\subsection{Semantic analysis of the posted content}

This module analyzes the textual content produced by users and provides a classification considering this perspective.
First, the posts associated with a specific user in the current snapshot  undergo a Natural Language Preprocessing pipeline: they are tokenized, stemmed, and concatenated. Then, for each obtained document, the Google's pretrained Word2Vec \cite{word2vec}
model\footnote{https://code.google.com/archive/p/word2vec/} is used to extract semantic features from the text. Such a model was trained on a corpus of part of Google News dataset (about 100 billion words), and is able to produce $300$-dimensional vectors for $3$ million words.
The embeddings of the words posted by a user are summed into a single feature vector that represents the user, exploiting the additive compositionality property of word embeddings \cite{mikolov2013distributed}.
More formally, the semantic representation ${s_u}^{(i)}$ of a user $u$ in a temporal network snapshot $i$ is computed as:
\begin{equation}
    {s_u}^{(i)} = \sum_{w \in {words^{(i)}(u)}} w2v(w)
\end{equation}
where ${words^{(i)}(u)}$ are the words present in the textual document associated with the user $u$ in the $i-$th snapshot.

Once the semantic representation for each user is obtained, two one-class autoencoders are trained. The first, which we refer to as ${A_R}^{(i)}$, is trained using the semantic representations associated with users labeled as \textit{risky}. The second, ${A_S}^{(i)}$, is trained with the semantic representations of the \textit{safe} users.
Then, the two autoencoders are fed with the user embedding vectors, and the reconstruction errors ${R_R}^{(i)}$ and ${R_S}^{(i)}$ are calculated as the mean squared error between the original vectors and the reconstructed ones.
The module outputs the two reconstruction errors ${R_R}^{(i)}$ and ${R_S}^{(i)}$, and a label ${L_{sem}}^{(i)}(u)$ computed as:
\begin{equation}
    {L_{sem}}^{(i)}(u)=
    \begin{cases}
    \text{0 if }{R_S}^{(i)} < {R_R}^{(i)} \\
    \text{1 otherwise}
    \end{cases}
\end{equation}
where 0 and 1 indicate  the \textit{safe} and \textit{risky} labels, respectively.

\subsection{Topological analysis of the relationship network}

\methodName{} represents the relationships among users in each snapshot with an adjacency matrix $A^{(i)} \in {\{0, 1\}}^{|N|\times|N|}$, where ${A^{(i)}_{hj}}=1$ if $(u_h, u_j) \in {E_T}^{(i)}$, and ${A^{(i)}_{hj}}=0$ otherwise.
$A^{(i)}$ is a highly dimensional and sparse matrix. Directly applying machine learning techniques to this kind of matrix is not optimal due to computational issues and the well-known \textit{curse of dimensionality}.
To address this problem, several techniques have been proposed for projecting a matrix into a lower-dimensional, more manageable, space. These methods are based on different strategies, such as Singular Value Decomposition \cite{DBLP:conf/cit/JaradatMJMA21}, Principal Component Analysis \cite{Salih}, or random walk-based embedding methods like DeepWalk \cite{deepwalk} and Node2Vec \cite{node2vec}.
We employ the latter approach, which merges random walks and the Word2Vec method to obtain a matrix $\hat{A}^{(i)} \in \mathbb{R}^{|N|\times k_r}$, where $k_r<<N$ is a user-defined parameter indicating the dimension of the resulting embedding.

The rows of the matrix $\hat{A}^{(i)}$ represent the embeddings associated with the corresponding users in the $i$-th snapshot and are used for training a classifier based on random forests.
At inference time the classifier trained on the $i^{th}$ snapshot will output, for the node to classify, a label ${L_{rel}}^{(i)}$, and a confidence score ${c_{rel}}^{(i)} \in [0, 1]$, that indicates the average purity of the leaf nodes of the forest trees in which the instance to classify falls: if the instance ends up in leaves where there is a strong prevalence of individuals of a class, the confidence will be high.

\subsection{Spatial analysis of the closeness among users}
The spatial analysis module of \methodName{} analyzes, for each snapshot, an undirected network depicting the spatial relationships among users. Specifically, the spatial network is represented as a weighted matrix $S^{(i)} \in \mathbb{R}^{|N| \times |N|}$, where ${S_{hj}}^{(i)} = closeness^{(i)}(u_h, u_j)$ corresponds to the spatial closeness between the user $u_h$ and the user $u_j$ in the $i$-th temporal snapshot.

To obtain the spatial closeness between two users, we take into account the mode of the geographic locations (intended as a latitude-longitude pair) associated with each user in the current snapshot. Consequently, each user may be associated with different locations throughout the temporal snapshots.  More formally, given two users $u_1, u_2$, their latitudes ${\phi_1}^{(i)}, {\phi_2}^{(i)}$ and their longitudes ${\lambda_1}^{(i)}, {\lambda_2}^{(i)}$, the distance $d^{(i)}(u_1, u_2)$ is computed as:
\begin{equation}
    d^{(i)}(u_1, u_2) = 2r \cdot arctan\frac{\sqrt{a^{(i)}(u_1, u_2)}}{\sqrt{1 - a^{(i)}(u_1, u_2)}}
    \end{equation}
\begin{equation}
a^{(i)}(u_1, u_2) = sin^2\left(\frac{{\phi_2}^{(i)} - {\phi_1}^{(i)}}{2}\right) + cos({\phi_1}^{(i)})\cdot cos({\phi_2}^{(i)})\cdot sin^2\left(\frac{{\lambda_2}^{(i)} - {\lambda_1}^{(i)}}{2}\right)
\end{equation}
where $r$ is the Earth radius ($\approx 6371km$).
After calculating the distances between each pair of available users in the considered (\textit{i}) snapshot, their mean ${\mu_d}^{(i)}$ and standard deviation ${\sigma_d}^{(i)}$ are determined, and they are used to compute z-score normalization $z^{(i)}(u_1, u_2)$ in $\mathcal{N}$(0,1). Then, the closeness value between two users is:
\begin{equation}
    {closeness^{(i)}(u_1, u_2)} = 
    \begin{cases}
    \frac{{z^{(i)}(u_1, u_2)}}{{min_z}^{(i)}}&\text{if $z^{(i)}(u_1, u_2)<0$} \\
    0 &\text{otherwise}
    \end{cases}
\end{equation}
Where ${min_z}^{(i)}$ is the minimum of the unnormalized distances among users for the current snapshot.
In other words, if two users are closer than the average, the closeness score will range in the interval $(0,1]$, otherwise, their closeness score will be set to zero (meaning that they are not connected in the spatial network of the analyzed snapshot).

As already done with the topological analysis of the relationship network, we use the Node2Vec algorithm to project each spatial adjacency matrix into a lower-dimensional vector space, obtaining a reduced weighted matrix $\hat{S}^{(i)} \in \mathbb{R}^{|N| \times k_s}$, where $k_s<<N$ is a user-defined parameter indicating the dimension of the resulting embedding.
Consequently, we train a node classifier model based on a decision tree that, at inference time, will output the label ${L_{spat}}^{(i)}$ and a confidence score ${c_{spat}}^{(i)}$ for each user.

\subsection{Model fusion}
In the final phase, the predictions made in the previous steps are combined to provide the classification about the risky class of users in a specific snapshot.
Specifically, considering a snapshot $i$, for each user $u$ the three previous components of \methodName{} return:
\begin{itemize}
    \item The risky and safe reconstruction errors of the autoencoders, namely, ${R_R}^{(i)}(u)$ and ${R_S}^{(i)}(u)$, and the label ${{L_{sem}}^{(i)}(u)}$ predicted by the semantic analysis of the posted content;
    \item The predicted label ${L_{rel}}^{(i)}(u)$ and the confidence ${c_{rel}}^{(i)}(u)$ predicted by topological analysis of the relationship network;
    \item The predicted label ${L_{spat}}^{(i)}(u)$ and the confidence ${c_{spat}}^{(i)}(u)$ predicted by spatial analysis of the closeness among users.
\end{itemize}
Following the stacked generalization approach, we concatenate this information into a vector that is fed to a Multi Layer Perceptron (MLP), which subsequently outputs the prediction $p(u)^{(i)}$: 
\begin{equation}
\resizebox{\hsize}{!}{${p(u)}^{(i)}=MLP(
{R_S}^{(i)}(u) 
\mathbin\Vert {R_R}^{(i)}(u) \mathbin\Vert {L_{sem}}^{(i)}(u) \mathbin\Vert {L_{rel}}^{(i)}(u) \mathbin\Vert {c_{rel}}^{(i)}(u) \mathbin\Vert {L_{spat}}^{(i)}(u) \mathbin\Vert {c_{spat}}^{(i)}(u)
)
$}
\nonumber
\end{equation}
\noindent
where $||$ is the concatenation operation. We recall that the models are independent from a snapshot to the other. They are trained separately and, at inference time, their final MLPs will produce $T-1$ predictions for the same user. 

\subsection{Voting phase}
After the model fusion phase is executed on each of the temporal snapshots, different labels may be obtained for the same user, indicating variations in their risk classification over time. To obtain a comprehensive risk label for each user, these single snapshot predictions need to be aggregated. This aggregation process aims at combining the labels from different time points, taking into account the temporal dynamics of the user's behavior.

The first and simplest strategy consists in using a majority voting mechanism, where the mode of the predicted labels is provided as the final output.
However, it is safe to assume that users change behavior gradually over time, hence information coming from closer temporal snapshots should be deemed as more important. If ${p(u)}^{(1)}, {p(u)}^{(2)}, ..., {p(u)}^{(T-1)}$ are the labels predicted for each snapshot, the final prediction will be:

\begin{equation}
    {p(u)}^{(T)} = max(score_{risky}, score_{safe})
\end{equation}

\noindent
where $score_{risky}$ and $score_{safe}$ are computed as follows:

\begin{equation}\label{eqt1}
    {score_{risky}} = \sum_{i=1}^{T-1} w^{(i)} \cdot \llbracket {p(u)}^{(i)} = risky \rrbracket
\end{equation}

\begin{equation}\label{eqt2}
    {score_{safe}} = \sum_{i=1}^{T-1} w^{(i)} \cdot \llbracket {p(u)}^{(i)} = safe \rrbracket
\end{equation}
and $w^{(i)}$ is the weight assigned to the $i^{th}$ snapshot by the weighting scheme. Specifically, we use and compare three different weighting schemes:
\begin{itemize}
    \item \textit{Uniform}, corresponding to the na\"ive method of majority voting. It gives the same weight to all the snapshots: $w^{(1)}=w^{(2)}=...=w^{(T-1)}=1$;
    \item \textit{Linear}: the importance of the predicted label increases linearly by time: $\forall i,\ 1\leq i \leq T - 1,\ w^{(i)}=i$;
    \item \textit{Quadratic}: the importance of the predicted labels increases quadratically by time: $\forall i,\ 1\leq i \leq T - 1,\ w^{(i)}=i^2$;
\end{itemize}

If a user is missing from a snapshot, the corresponding term in Equations \ref{eqt1}, \ref{eqt2} is ignored. Note that the second and the third weighting schema implicitly introduce some form of "forgetting", meaning that the final resulting model gradually loses information about past users' activities.

\subsection{Graph smoothing}
\label{sec:smoothing}
The temporal snapshots of \methodName{} are partially overlapping, meaning that part of the content that a user posted in a snapshot also appears in the next snapshot. This allows the textual embedding associated with a user to accurately capture the gradual evolution in the way users think and express themselves.
On the other hand, the topological and spatial embeddings can vary significantly from snapshot to snapshot for each user, even if the neighborhood remains largely unchanged, because the Node2Vec computation considers other nodes in the network. 
This aspect can negatively impact the performance of the classifiers that will provide a prediction for the topological and spatial analysis. 

To address this issue, \methodName{} adopts a \textit{graph smoothing} method, which, for each user $u$, incorporates information about the topological (or spatial) embedding from the previous snapshots into the current one.
This is done by computing the node representation in the current snapshot and summing it to the embeddings obtained for the same user in the previous snapshots.
It results in a \textit{smoothed} representation, which is then used as input for the classifiers.
Formally:
\begin{equation}
    smooth(u^{(i)}) = \sum_{t \in \{1, 2,...i-1\} \text{ s.t. } u \in N^{(t)}} n2v^{(t)}(u)^{(t)} 
\end{equation}
Consequently, if there is a snapshot $t$ such that $u \not\in N^{(t)}$, it is skipped from the computation of the user embedding.

\section{Experiments}\label{sec4}
In this section, we first describe the dataset and the experimental setup. Then, we show and discuss the obtained results.

\subsection{Dataset}
The considered dataset is extracted from Twitter through a compliant crawling system. The crawling procedure consisted in extracting tweets posted in the span that ranges from June 2014 to July 2018, on the basis of a set of keywords belonging to the homeland security field. The keywords were determined within the research activities of the EU project CounteR (https://counter-project.eu/). The dataset contains 117,570 users and approximately $118$ millions  tweets posted. Each tweet is associated with a timestamp, the coordinates (latitude and longitude) of the location from where it was posted and a sentiment score computed with the CoreNLP toolkit (https://stanfordnlp.github.io/CoreNLP/).

\begin{figure}[tb]
    \centering
    \includegraphics[width=0.8\textwidth]{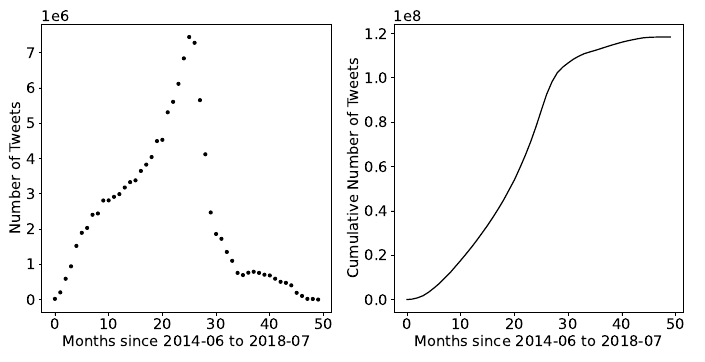}
    \caption{Distribution and cumulative distribution of the number of tweets per month in the considered dataset.}
    \label{fig:graphics}
\end{figure}

Figure \ref{fig:graphics} shows the distribution of tweets over time. As we can observe, the distribution is not uniform, with the number of tweets varying significantly across different periods. Consequently, splitting the dataset considering an equal temporal width would result in snapshots containing a highly varying number of tweets and users. 
Therefore, we adjusted the splitting timestamps to ensure that each snapshot contains a similar number of tweets and exhibit a partial overlap.

We performed two different dataset splittings to create two experimental scenarios: $\mathcal{D}_5$, a 5-snapshots split, and $\mathcal{D}_{10}$, a 10-snapshots split, as shown in Table \ref{tab:table_distr}. Since the last snapshot in each split is used for testing, we removed from it all the users who do not appear in any of the previous snapshots.

In view of an practical application of the framework in a real-world scenario, a crucial decision point is determining the optimal timing for performing a new split. Our suggestion is to split when the amount of gathered tweets and users approximates that of the previous snapshots. However, since the models trained on the snapshots are independent from each other, there are no strict constraints. A user might deem appropriate to perform a new split when a significant social or political event occurs, in order to keep it into account for future user evaluations.

\begin{table}[tb]
\centering
\begin{tabular}{|c|l|l|r|r|r|r|}
\hline
\textit{\textbf{Dataset}}               & \textit{\textbf{From}}       & \textit{\textbf{To}}         & \textit{\textbf{\#Users}} & 
\textit{\textbf{\#Risky}} & \textit{\textbf{\#Safe}} & \textit{\textbf{\#Tweets}}  \\ \hline
\multirow{5}{*}{$\mathcal{D}_5$} & 2014-06-22 & 2015-07-31 & 4,698 & 2,162 & 2,536 & 4,922,590 \\
 & 2015-06-01 & 2016-02-16 & 5,367 & 2,470 & 2,897 & 5,610,190 \\
 & 2015-12-18 & 2016-06-30 & 5,835 & 2,623 & 3,212 & 6,788,843 \\
 & 2016-05-01 & 2016-10-11 & 5,758 & 2,417 & 3,339 & 8,646,021 \\
 & 2016-08-12 & 2018-07-11 & 4,739 & 2,030 & 2,709 & 8,069,484 \\ \hline
\multirow{10}{*}{$\mathcal{D}_{10}$} & 2014-06-22 & 2015-03-28 & 5,334 & 2,452 & 2,882 & 3,647,227 \\
 & 2015-01-27 & 2015-07-31 & 5,972 & 2,707 & 3,265 & 4,055,008 \\
 & 2015-06-01 & 2015-11-17 & 6,408 & 2,913 & 3,495 & 4,330,119 \\
 & 2015-09-18 & 2016-02-16 & 6,782 & 3,088 & 3,694 & 4,651,140 \\
 & 2015-12-18 & 2016-04-30 & 7,266 & 3,261 & 4,005 & 5,249,484 \\
 & 2016-03-01 & 2016-06-30 & 7,428 & 3,202 & 4,226 & 6,177,391 \\
 & 2016-05-01 & 2016-08-21 & 7,433 & 3,120 & 4,313 & 7,406,486 \\
 & 2016-06-22 & 2016-10-11 & 7,194 & 2,981 & 4,213 & 7,967,931 \\
 & 2016-08-12 & 2017-01-28 & 6,151 & 2,583 & 3,568 & 6,454,669 \\
 & 2016-11-29 & 2018-07-11 & 2,410 & 1,094 & 1,316 & 403,916   \\ \hline
\end{tabular}
\label{tab:table_distr}
\caption{The two splittings performed on the dataset}
\end{table}

\subsection{Experimental setup}
We compared the performance of our proposed method on both the dataset variants against those obtained by the \textit{static} version of \textbf{SAIRUS} \cite{sairus}, which does not take into account the temporal evolution of users.
Specifically, the competitor was trained on the dataset obtained by merging information in the first $T-1$ snapshots and was then tested on the last snapshot.

All the three weighting schemes for the voting phase were considered in our experiments, leading to three variants of T-SAIRUS, namely \textbf{T-SAIRUS(U)}, based on uniform weighting, \textbf{T-SAIRUS(L)}, based on the linear weighting, and \textbf{T-SAIRUS(Q)}, based on the quadratic weighting. Moreover, to evaluate the effect of adopting the graph smoothing principle, we show how the performance of \methodName{} changes when this principle is not applied to the topological and spatial embeddings.

As regards the embeddings dimensionality, taking into account the best results obtained by SAIRUS \cite{sairus}, we set $k_r$ and $k_s$ to 128, while the Word2Vec embedding size was set to 300. The random forests for the node classification have 100 estimators, with maximum depth parameter set to 5. As evaluation measures, we considered \textit{precision}, \textit{recall}, \textit{accuracy}, and \textit{F1-score}, which are computed for the entire user set as well as for individual classes.

The experiments were performed on a server equipped with a NVIDIA GeForce Titan X GPU and 64 GB of RAM. The \methodName{} code is available on Github\footnote{https://github.com/itsfrank98/tsairus}.

\subsection{Results and Discussion}

\begin{table}[!b]
\begin{tabular}{|c|c|cccc|ccc|ccc|}
\hline
\multirow{2}{*}{\textbf{\textit{Method}}} & \textit{\textbf{Graph}} & \multicolumn{4}{c|}{\textbf{\textit{All users}}} & \multicolumn{3}{c|}{\textbf{\textit{Safe}}} & \multicolumn{3}{c|}{\textbf{\textit{Risky}}} \\ \cline{3-12} 
\multicolumn{1}{|c|}{\textbf{}} & \textbf{\textit{Smoothing}} & \multicolumn{1}{c}{\textbf{\textit{Prec}}} & \multicolumn{1}{c}{\textbf{\textit{Rec}}} & \multicolumn{1}{c}{\textbf{\textit{F1}}} & \textbf{\textit{Acc}} & \multicolumn{1}{c}{\textbf{\textit{Prec}}} & \multicolumn{1}{c}{\textbf{\textit{Rec}}} & \textbf{F1} & \multicolumn{1}{c}{\textbf{\textit{Prec}}} & \multicolumn{1}{c}{\textbf{\textit{Rec}}} & \textbf{\textit{F1}} \\ \hline \hline
\multicolumn{1}{|c|}{SAIRUS \cite{sairus}} & \multirow{4}{*}{No} & \multicolumn{1}{c}{0.69} & \multicolumn{1}{c}{0.54} & \multicolumn{1}{c}{0.61} & 0.48 & \multicolumn{1}{c}{0.45} & \multicolumn{1}{c}{\textbf{0.99}} & 0.62 & \multicolumn{1}{c}{\textbf{0.93}} & \multicolumn{1}{c}{0.08} & 0.14 \\ \cline{1-1} \cline{3-12} 
\multicolumn{1}{|c|}{T-SAIRUS (U)} &  & \multicolumn{1}{c}{\textbf{0.77}} & \multicolumn{1}{c}{0.55} & \multicolumn{1}{c}{0.64} & 0.61 & \multicolumn{1}{c}{\textbf{0.95}} & \multicolumn{1}{c}{0.11} & 0.20 & \multicolumn{1}{c}{0.59} & \multicolumn{1}{c}{\textbf{1.00}} & 0.74 \\ \cline{1-1} \cline{3-12} 
\multicolumn{1}{|c|}{T-SAIRUS (L)} &  & \multicolumn{1}{c}{\textbf{0.77}} & \multicolumn{1}{c}{0.55} & \multicolumn{1}{c}{0.64} & 0.61 & \multicolumn{1}{c}{\textbf{0.95}} & \multicolumn{1}{c}{0.11} & 0.20 & \multicolumn{1}{c}{0.59} & \multicolumn{1}{c}{\textbf{1.00}} & 0.74 \\ \cline{1-1} \cline{3-12} 
\multicolumn{1}{|c|}{T-SAIRUS (Q)} &  & \multicolumn{1}{c}{0.76} & \multicolumn{1}{c}{0.53} & \multicolumn{1}{c}{0.62} & 0.59 & \multicolumn{1}{c}{0.93} & \multicolumn{1}{c}{0.07} & 0.13 & \multicolumn{1}{c}{0.58} & \multicolumn{1}{c}{\textbf{1.00}} & 0.73 \\ \hline \hline
\multicolumn{1}{|c|}{T-SAIRUS (U)} & \multirow{3}{*}{Yes} & \multicolumn{1}{c}{0.76} & \multicolumn{1}{c}{\textbf{0.76}} & \multicolumn{1}{c}{\textbf{0.76}} & \textbf{0.75} & \multicolumn{1}{c}{0.66} & \multicolumn{1}{c}{0.87} & \textbf{0.75} & \multicolumn{1}{c}{0.87} & \multicolumn{1}{c}{0.66} & \textbf{0.75} \\ \cline{1-1} \cline{3-12} 
\multicolumn{1}{|c|}{T-SAIRUS (L)} &  & \multicolumn{1}{c}{0.74} & \multicolumn{1}{c}{0.73} & \multicolumn{1}{c}{0.73} & 0.71 & \multicolumn{1}{c}{0.62} & \multicolumn{1}{c}{0.89} & 0.73 & \multicolumn{1}{c}{0.87} & \multicolumn{1}{c}{0.58} & 0.70 \\ \cline{1-1} \cline{3-12} 
\multicolumn{1}{|c|}{T-SAIRUS (Q)} &  & \multicolumn{1}{c}{0.75} & \multicolumn{1}{c}{0.75} & \multicolumn{1}{c}{0.75} & 0.73 & \multicolumn{1}{c}{0.63} & \multicolumn{1}{c}{0.88} & 0.74 & \multicolumn{1}{c}{0.87} & \multicolumn{1}{c}{0.61} & 0.72 \\ \hline
\end{tabular}
\caption{Results on the $\mathcal{D}_5$ dataset}
\label{tab:d5_results}
\end{table}

Table \ref{tab:d5_results} shows the results achieved on the $\mathcal{D}_{5}$ dataset, with the best performance for each metric highlighted in bold. \methodName{} significantly outperforms the competitor when graph smoothing is adopted, with the uniform voting scheme yielding the best overall results on the dataset. However, it can be observed that the competitor SAIRUS outperforms the variant of our method without graph smoothing. Notably, SAIRUS achieves the highest recall on the \textit{safe} class and precision on the \textit{risky} class, primarily because it predicts the \textit{safe} label to most of the testing instances. Considering that the dataset is quite balanced, this is not an expected behavior, which can however be due to the fact that SAIRUS observes the training data altogether, without considering the temporal dimension. This limitation probably makes it unable to properly detect  users who acted safely in the past that are possibly evolving towards being risky. In contrast, \methodName{} appears to  be more robust to these situations. 

Table \ref{tab:d10_results} shows the results achieved on the $\mathcal{D}_{10}$ dataset. 
In this case, our method with the uniform weighting scheme and without graph smoothing proves to be the best among the variants we evaluated. However, there is no clear winner in the comparison between our method and the competitor, SAIRUS. \methodName{} obtains overall better results on the \textit{risky} class, i.e., in terms of F1-score, and achieves a higher recall in identifying \textit{risky} users. This indicates that \methodName{} is more effective at recognizing dangerous users and is less prone to produce false negatives, which is preferable given that the primary goal is to detect dangerous users. Additionally, we stress that \methodName{} is more computationally efficient than SAIRUS: when a new node is added, \methodName{} only needs to be trained on the last snapshot, whereas SAIRUS requires a complete retraining.

\begin{table}[!t]
\begin{tabular}{|c|c|cccc|ccc|ccc|}
\hline
\multirow{2}{*}{\textbf{\textit{Method}}} & \textit{\textbf{Graph}} & \multicolumn{4}{c|}{\textbf{\textit{All users}}} & \multicolumn{3}{c|}{\textbf{\textit{Safe}}} & \multicolumn{3}{c|}{\textbf{\textit{Risky}}} \\ \cline{3-12} 
\multicolumn{1}{|c|}{} & \textbf{\textit{Smoothing}} & \textbf{\textit{Prec}} & \textbf{\textit{Rec}} & \textbf{\textit{F1}} & \textbf{\textit{Acc}} & \textbf{\textit{Prec}} & \textbf{\textit{Rec}} & \textbf{\textit{F1}} & \textbf{\textit{Prec}} & \textbf{\textit{Rec}} & \textbf{\textit{F1}} \\ \hline \hline
\multicolumn{1}{|c|}{SAIRUS \cite{sairus}} & \multirow{4}{*}{No} & \textbf{0.75} & 0.63 & \textbf{0.68} & 0.63 & 0.57 & \textbf{0.98} & \textbf{0.72} & \textbf{0.92} & 0.29 & 0.44 \\ \cline{1-1} \cline{3-12} 
\multicolumn{1}{|c|}{T-SAIRUS (U)} &  & 0.67 & \textbf{0.64} & 0.65 & \textbf{0.64} & \textbf{0.60} & 0.83 & 0.70 & 0.74 & \textbf{0.46} & \textbf{0.56} \\ \cline{1-1} \cline{3-12} 
\multicolumn{1}{|c|}{T-SAIRUS (L)} &  & 0.63 & 0.59 & 0.60 & 0.59 & 0.55 & 0.87 & 0.67 & 0.71 & 0.32 & 0.44 \\ \cline{1-1} \cline{3-12} 
\multicolumn{1}{|c|}{T-SAIRUS (Q)} &  & 0.51 & 0.50 & 0.50 & 0.50 & 0.49 & 0.89 & 0.64 & 0.52 & 0.12 & 0.19 \\ \hline \hline
\multicolumn{1}{|c|}{T-SAIRUS (U)} & \multirow{3}{*}{Yes} & 0.41 & 0.43 & 0.42 & 0.43 & 0.45 & 0.69 & 0.54 & 0.38 & 0.18 & 0.25 \\ \cline{1-1} \cline{3-12} 
\multicolumn{1}{|c|}{T-SAIRUS (L)} &  & 0.31 & 0.39 & 0.35 & 0.38 & 0.42 & 0.70 & 0.53 & 0.20 & 0.07 & 0.11 \\ \cline{1-1} \cline{3-12} 
\multicolumn{1}{|c|}{T-SAIRUS (Q)} &  & 0.29 & 0.38 & 0.33 & 0.37 & 0.42 & 0.70 & 0.52 & 0.17 & 0.06 & 0.09 \\ \hline
\end{tabular}
\caption{Results on the $\mathcal{D}_{10}$ dataset}
\label{tab:d10_results}
\end{table}

In contrast to the previous scenario, graph smoothing appears to be detrimental to the results. A possible explanation is that graph smoothing is performed by summing node embeddings from all previous snapshots, which have an equal weight, independently on the temporal closeness to the testing snapshot.

\vspace{-10pt}
\section{Conclusions}
\label{sec:conclusions}
\vspace{-10pt}
In this paper, we proposed \methodName{}, a multi-view learning approach for user classification in social networks that considers the temporal evolution of people's thoughts, social connections, and spatial closeness. This is achieved by dividing the network into temporal snapshots and learning a dedicated model for each snapshot. When classifying a user, the predictions made by each model are combined to provide a temporally-aware classification. The proposed strategy is also computationally efficient, as it does not require complete retraining when a new user (or a new batch of users) is added to the network: only the model trained for the snapshot in which the user is (users are) added needs to be retrained, allowing to reuse models trained from the previous snapshots.

\methodName{} was evaluated on a real-world Twitter dataset, which was split into 5 and 10 snapshots. The achieved performances were compared to those obtained by a state-of-the-art competitor that does not consider the temporal dimension, proving the effectiveness of our approach.

For future developments, we plan to enhance the graph smoothing strategy. Currently, the graph smoothing used in our experiments assigns the same weight to all snapshots, which negatively impacted the tests on the 10-snapshot dataset. We believe that a different strategy, prioritizing more recent snapshots, would be beneficial to the system. 
Additionally, other node embedding strategies could be explored: feature-based methods, in particular, might better capture the user evolution compared to random walk-based strategies.

\begin{credits}
\subsubsection{\ackname} 
The authors acknowledge the support of the European Commission through
the H2020 Project “CounteR - Privacy-First Situational Awareness Platform
for Violent Terrorism and Crime Prediction, Counter Radicalisation and Citizen
Protection” (Grant N. 101021607). 
This work was partially supported by the project FAIR - Future AI Research (PE00000013), spoke 6 - Symbiotic AI, under the NRRP MUR program funded by the NextGenerationEU.
\end{credits}

\bibliographystyle{splncs04}
\bibliography{samplebib}

\end{document}